\documentclass[pra,twocolumn,showpacs,letterpaper]{revtex4}
\usepackage[dvips]{graphics,graphicx}
\usepackage{amssymb}
\usepackage{bm}
\usepackage{amsmath}

\newcommand{\OH}{{\hat{H}}}
\newcommand{\OW}{{\hat{W}}}
\newcommand{\OP}{{\hat{\ef}}}
\newcommand{\Oa}{{\hat{a}}}
\newcommand{\Oaa}{{\hat{a}^\dagger}}
\newcommand{\ON}{{\hat{N}}}
\newcommand{\On}{{\hat{n}}}
\newcommand{\ef}{{\Phi}}
\newcommand{\ket}[1]{|#1\rangle}
\newcommand{\scal}[2]{\ensuremath{\langle #1 | #2 \rangle}}
\newcommand{\diff}[2]{\frac{\partial #1}{\partial #2}}
\newcommand{\Diff}[2]{\frac{{\rm d} #1}{{\rm d} #2}}
\renewcommand{\mod}[0]{\,\text{mod}\,}
\newcommand{\N}{\mbox{$\mbox{I}\hspace{-2.2pt}\mbox{N}$}}

\begin{document}
\title{Semiclassical approach to Bose-Einstein condensates in a triple well potential}
\author{S. Mossmann}
\email{mossmann@fis.unam.mx}
\author{C. Jung}
\affiliation{Centro de Ciencias F\'{\i}sicas UNAM, 62251 Cuernavaca, M\'exico}
\date{\today }

\begin{abstract}
We present a new approach for the analysis of Bose-Einstein condensates in a few mode approximation. This method has already been used to successfully analyze the vibrational modes in various molecular systems and offers a new perspective on the dynamics in many particle bosonic systems. 
We discuss a system consisting of a Bose-Einstein condensate in a triple well potential. Such systems correspond to classical Hamiltonian systems with three degrees of freedom. The semiclassical approach allows a simple visualization of the eigenstates of the quantum system referring to the underlying classical dynamics. From this classification we can read off  the dynamical properties of the eigenstates such as particle exchange between the wells and entanglement without further calculations. In addition, this approach offers new insights into the validity of the mean-field description of the many particle system by the Gross-Pitaevskii equation, since we make use of exactly this correspondence in our semiclassical analysis.
We choose a three mode system in order to visualize it easily and, moreover, to have a sufficiently interesting structure, although the method can also be extended to higher dimensional systems.
\end{abstract}
\maketitle


\section{Introduction}
Bose-Einstein condensates form one of the main topics of research at the moment. One reason for this enormous interest is the fact that they combine concepts and techniques from different areas of physics, such as quantum optics, condensed matter physics, molecular physics and quantum chaos.
On the experimental side, there has been a remarkable progress in confining and manipulating Bose-Einstein condensates \cite{Albi05,Schu05a}, which has stimulated the theoretical research in the area.

There has been a large number of previous studies analyzing the dynamics of Bose-Einstein condensates in a double well potential using a mean-field approach, the Gross-Pitaevskii equation. In \cite{Smer97} Smerzi {\it et~al.}\ discussed the occurrence of macroscopic self-trapping within one well.
The behavior of the system during an adiabatic change of parameters was studied in \cite{06zener_bec,Liu03,Wu03} and generalizations of the linear two level crossing scenarios and the Landau-Zener formula were analyzed. Another line of investigation considers the mesoscopic regime in which the quantum and the classical, i.e.\ mean-field descriptions overlap and therefore semiclassical techniques can be used to study the system \cite{Milb97,Smer00,Angl01a,Angl01b,Thom03,Mahm05}.

In this paper we present a semiclassical technique to analyze the spectral properties of a Bose-Einstein condensate. This method has already been used to describe the vibrational spectra of molecules \cite{Sibe96,Jaco99,Jung02,Jung04} and provides an intuitive picture for the at first sight uninterpretable spectra. In this sense, the dynamics of a Bose-Einstein condensate in a triple well potential is analogous to the vibrations of a tri-atomic molecule.
Using the geometrical language of classical mechanics to describe the quantum system, we introduce a simple semiclassical method of visualization and classification of the quantum eigenstates which allows a characterization of the dynamics of the system. Furthermore, we investigate how far the correspondence between the mean-field system and the quantum many body system can be extended when the number of particles decreases.

For our studies we consider a Bose-Einstein condensate in a triple well potential, since the technique easily allows us to go beyond the standard double well potential analysis. A triple well potential has a much richer structure \cite{05level3,Nemo00,Thom03,Buon03,Fran03,Buon04} and the power of the method can be shown without loss of clarity, still allowing a direct visualization of all relevant structures.
\section{The model}\label{sec-model}
In the following analysis we consider a system consisting of bosonic particles in an external periodic potential $V(\vec{r})=V(\vec{r}+\vec{r}_{\vec{l}})$ with $\vec{r}_{\vec{l}}= l_1d_1\vec{e}_1+l_2d_2\vec{e}_2+l_3d_3\vec{e}_3$, $l_k\in\N$ and $d_k\in{\mathbf{R}}$. If a weak two-particle point-like interaction is assumed, then the Hamiltonian in second quantization can be written as
\begin{align}
\OH =& \int d^3r\; \OP^\dagger(\vec{r})\Bigl[-\frac{\hbar^2}{2m}\Delta + V(\vec{r})\Bigr]\OP(\vec{r}) \nonumber \\
&+ \frac{g}{2} \int d^3r\; \OP^\dagger(\vec{r})\,\OP^\dagger(\vec{r})\,\OP(\vec{r})\,\OP(\vec{r})\,.
\end{align}
Here, $m$ is the particle mass, $g=4\pi a_s\hbar^2/m$ is the coupling constant describing two-body interactions and $a_s$ is the s-wave scattering length.
For a repulsive interaction, $g$ is positive while for an attractive interaction $g$ takes a negative value.
For the rest of the paper we choose scaled units with $\hbar=m=1$. 
The field operator $\OP(\vec{r})$ can be expanded in terms of bosonic annihilation operators,
\begin{equation}
\OP(\vec{r}) = \sum_{n,m}{\phi_{n,m}}(\vec{r})\,\Oa_{n,m}\,,
\end{equation}
where we assume that the basis functions $\{\phi_{n,m}\}$ of the one-particle Hilbert space are exponentially localized in space and real, as is the case for the Wannier functions \cite{Kohn59}. The index $n$ describes basis functions in different wells and we will take into account only three different wells in order to model the three well potential. The second index $m$ labels the excited states within a single well. 
Assuming Bose-Einstein condensates, we can restrict ourselves to the lowest energy state $m =1$ and neglect higher excited states (see also \cite{Milb97} for a careful discussion of this topic for a two well potential).
Experimentally such a system was realized in \cite{Albi05} for a two well potential but the technique can in principle also be extended to three wells.

Expanding the Hamiltonian in this basis and neglecting fourth order terms in the creation and annihilation operators from different basis functions (modes)  yields the well-known Bose-Hubbard Hamiltonian \cite{Fish89b} restricted to three wells. So, the Hamiltonian can be written in a symmetrized form as
\begin{equation}\label{eq-Hqm}
\OH = \OH_0 + \OW
\end{equation}
with
\begin{align}
\OH_0 =&\;\omega_1\,\frac{\Oaa_1\Oa_1+\Oa_1\Oaa_1}{2}
+\omega_2\,\frac{\Oaa_2\Oa_2+\Oa_2\Oaa_2}{2}\nonumber\\&+
\omega_3\,\frac{\Oaa_3\Oa_3+\Oa_3\Oaa_3}{2}
+x_{1}\,\biggl(\frac{\Oaa_1\Oa_1+\Oa_1\Oaa_1}{2}\biggl)^2\nonumber\\
&+ x_{2}\,\biggl(\frac{\Oaa_2\Oa_2+\Oa_2\Oaa_2}{2}\biggl)^2
+x_{3}\,\biggl(\frac{\Oaa_3\Oa_3+\Oa_3\Oaa_3}{2}\biggl)^2\,,\\
\OW =&\; -\frac{k_{12}}{2}\,(\Oaa_1\Oa_2+\Oaa_2\Oa_1)
-\frac{k_{23}}{2}\,(\Oaa_2\Oa_3+\Oaa_3\Oa_2)\,.\label{eq-Hqmb}
\end{align}
Here we neglect a constant energy shift. For convenience, we will choose the nonlinear interaction strengths $x_j$ equal for each well in the following sections which is also in accordance with experimental realizations.
Such Hamiltonians have already been studied in great detail for the more restrictive two mode model (e.g.\ in \cite{Spek99, Angl01b, Mahm05}). 

The Hamiltonian commutes with the particle number operator $\ON = \On_1+\On_2+\On_3$ which expresses the conservation of the total number of particles. The symmetrized form is more convenient when considering the semiclassical limit, as will become clear in the next paragraph.
Hamiltonians of this kind have been used in molecular physics in order to describe and assign vibrational spectra \cite{Jung04,Sibe96}. In the molecular case they describe all kinds of vibrational degrees of freedom like stretches, bends, torsions etc.\ and include various resonant interactions corresponding to different simple rational ratios between the frequencies. The conserved particle number in our case of Eq.~(\ref{eq-Hqm}) corresponds to the polyad-type conserved quantities in the molecular systems.

For the case of 30 particles considered in the following, it is an easy numerical task to diagonalize the Hamiltonian matrix and thus solve the problem. However, one cannot understand the underlying structure of this system from numerical values alone. The aim of this paper is to present a method which allows an easy visual characterization of the eigenvectors of the Hamiltonian, using the close correspondence with the classical system.
\subsection{The classical system}
Essential for our semiclassical classification and assignment of quantum states
is a comparison between the quantum states and the corresponding classical 
dynamics. To this end, the first step is the construction of the classical
Hamiltonian function, which corresponds to the quantum Hamiltonian given in
Eqs.~(\ref{eq-Hqm})--(\ref{eq-Hqmb}). This is done by Heisenberg's substitution rules \cite{Heis25}
\begin{equation}\label{eq-aIrel}
\Oa_k\rightarrow \sqrt{I_k}e^{i\varphi},\quad
\Oaa_k\rightarrow \sqrt{I_k}e^{-i\varphi}\,.
\end{equation}
There are two different lines of argumentation for this substitution.
First, it is exact for the harmonic oscillator where the well known classical
Hamiltonian $\omega I$ is obtained by the replacement of the symmetrized 
product of an annihilation and a creation operator by the classical action.
This implies the correspondence
\begin{equation}\label{eq-Inrel1}
I \longleftrightarrow n+\frac{1}{2}
\end{equation}
between the classical action $I$ and the quantum number $n$ of the oscillator
($I$ is here measured in 
units of $\hbar$). This correspondence of Eq.~(\ref{eq-Inrel1}) is also a result of the 
application of the semiclassical Bohr-Sommerfeld quantization rules to the 
harmonic oscillator. In
more general cases we have to generalize the Bohr-Sommerfeld method to the
EBK quantization. Then the argument holds for any bound system
of any number of degrees of freedom as long as the system is close to 
integrable  (for general background information on
semiclassics see \cite{Brac97}). In general, the semiclassical methods give 
results correct in the lowest two orders in $\hbar$ (orders 0 and 1) and
cause errors of order $\hbar^2$.
The application of the substitution rules of Eq.~(\ref{eq-aIrel}) to the quantum Hamiltonian
of Eqs.~(\ref{eq-Hqm})--(\ref{eq-Hqmb}) gives
\begin{align}\label{eq-Hcl}
&H(\varphi_1,\varphi_2,\varphi_3,I_1,I_2,I_3) \nonumber\\&= H_0(I_1,I_2,I_3)+W(\varphi_1,\varphi_2,\varphi_3,I_1,I_2,I_3)
\nonumber\\&
=\omega_1\,I_1+\omega_2\,I_2+
\omega_3\,I_3+x_{1}\,I_1^2+x_{2}\,I_2^2+x_{3}\,I_3^2\\&\quad
-k_{12}\sqrt{I_1I_2}\,\cos(\varphi_1-\varphi_2)
-k_{23}\sqrt{I_2I_3}\,\cos(\varphi_2-\varphi_3)\,.\nonumber
\end{align}
A Hamiltonian for the same system but expanded in another basis was analyzed in \cite{Thom03}.
This function can be interpreted as the Hamiltonian of a classical system of three coupled anharmonic oscillators described in action-angle variables $\varphi_k\in[0,2\pi)$ and $I_k>0$, where $k=1,2,3$. 
As a method to construct the corresponding 
classical Hamiltonian, the substitution rules of Eq.~(\ref{eq-aIrel}) always give the correct 
result since in this direction (quantum $\rightarrow$ classical) the
correspondence is unique whenever it exist at all, in contrast to the other
direction (classical $\rightarrow$ quantum) with its notorious $\hbar^2$
problems.
At high excitation (large quantum numbers)
there is a second argument for the semiclassical correspondence. The
application of a creation or annihilation operator to a number
state $\ket{n}$ has the effect
\begin{equation}
\Oa\, \ket{n} = \sqrt{n}\, \ket{n-1},\quad \Oaa\, \ket{n} = \sqrt{n+1}\, \ket{n+1}\,.
\end{equation}
In the limit of a large quantum number $n$, the difference between $n$ and
$n+1$ or $n-1$ is irrelevant in the square roots as well as in the states
and the operators can simply be replaced by multiplication with the
number $\sqrt{n}$. This argument
holds for condensates where a large number of particles goes into a
superfluid state which is well described by a mean-field limit. This is
in line with the standard argument of semiclassical behavior in the limit
of large quantum numbers. Interestingly, for systems of coupled anharmonic 
oscillators the semiclassical treatment is very good also for low excitation
numbers. In this limit, we approach the integrable 
harmonic limit where the Bohr-Sommerfeld treatment gives the correct result.
The experience with molecular systems of the structure of 
Eqs.~(\ref{eq-Hqm})--(\ref{eq-Hqmb}) shows
that a semiclassical treatment of such systems is globally quite good in
most cases.

Accordingly, we base our method of semiclassical assignment on this argument.
Semiclassical arguments will be
used later first to convert the eigenstates of the many-body Hamiltonian (\ref{eq-Hqm}) into 
wave functions on the toroidal configuration space and second to compare these
functions with important structures seen in the classical dynamics.

The integrable part $H_0$ of the Hamiltonian, which does not contain interactions 
between the three oscillators, leaves all actions unchanged.
In contrast, $W$ changes the values of the 
actions (particles in the wells) because of its dependence on angles 
and introduces interactions
between the three oscillators. In this sense we call in the following $W$ 
the interaction part of the Hamiltonian. In the picture of particles in
the triple well, $W$ describes tunneling terms between the various wells.

The Poisson bracket between $H$ and the observable 
\begin{equation}\label{eq-Kcl}
K=I_1+I_2+I_3\,,
\end{equation} 
the total action, is equal to zero, which corresponds to the quantum mechanically conserved number of particles. Note that the numerical value of $K$ differs by $3\cdot 1/2$ from the value of $N$ because of the zero point actions. The symmetry $\{H,K\}=0$ can be used to reduce the number of degrees of freedom from three to two by a canonical transformation. Using the generating function 
\begin{multline}
G(\varphi_1,\varphi_2,\varphi_3,J_1,J_2,K) \\= J_1(\varphi_1-\varphi_2) + J_2(\varphi_3-\varphi_2) + K \varphi_2
\end{multline}
of the old angles $(\varphi_1,\varphi_2,\varphi_3)$ and the new actions $(J_1,J_2,K)$ results in the transformations (together with Eq.~(\ref{eq-Kcl}))
\begin{eqnarray}
&\psi_1 = \varphi_1-\varphi_2\,,\quad\psi_2 = \varphi_3-\varphi_2\,,\quad \vartheta=\varphi_2\,,\nonumber\\
&I_1 = J_1\,,\quad I_3 = J_2\,,\label{eq-vartrans}
\end{eqnarray}
where $(\psi_1,\psi_2,\theta)$ are the new angles conjugate to $(J_1,J_2,K)$.

The Hamiltonian in the new coordinates is given by
\begin{align}\label{eq-Hclred}
H =& \;\omega_1\,J_1+\omega_2\,(K-J_1-J_2)+
\omega_3\,J_2\nonumber\\&
+x_{1}\,J_1^2+x_{2}\,(K-J_1-J_2)^2+x_{3}\,J_2^2\nonumber\\&
-k_{12}\sqrt{J_1(K-J_1-J_2)}\,\cos\psi_1\nonumber\\&
-k_{23}\sqrt{J_2(K-J_1-J_2)}\,\cos\psi_2\,,
\end{align}
with corresponding equations of motions
\begin{align}
\dot{\psi}_1 &=(\omega_1+2x_1J_1)-(\omega_2+2x_2(K-J_1-J_2))\nonumber\\
&\quad -\frac{k_{12}}{2}\Biggl[\,\sqrt{\frac{K-J_1-J_2}{J_1}}-\sqrt{\frac{J_1}{K-J_1-J_2}} \;\Biggr]\cos\psi_1\nonumber\\
&\quad + \frac{k_{23}}{2}\sqrt{\frac{J_2}{K-J_1-J_2}}\cos\psi_2\,,\label{eq-diffhama}\\
\dot{\psi}_2 &=(\omega_3+2x_3J_2)-(\omega_2+2x_2(K-J_1-J_2))\nonumber\\
&\quad -\frac{k_{23}}{2}\Biggl[\,\sqrt{\frac{K-J_1-J_2}{J_2}}-\sqrt{\frac{J_2}{K-J_1-J_2}} \;\Biggr]\cos\psi_2\nonumber\\
&\quad + \frac{k_{12}}{2}\sqrt{\frac{J_1}{K-J_1-J_2}}\cos\psi_1\,,\label{eq-diffhamb}\\
\dot{J}_1 &=-k_{12}\,\sqrt{J_1(K-J_1-J_2)}\,\sin\psi_1\,,\label{eq-diffhamc}\\
\dot{J}_2 &=-k_{23}\,\sqrt{J_2(K-J_1-J_2)}\,\sin\psi_2\,.\label{eq-diffhamd}
\end{align}
The classical configuration space is a two dimensional torus spanned by the two angles  $\psi_1$ and $\psi_2$. In order to compare the classical and the quantum system we have to represent the states as wave functions on the classical configuration space. The way to do this will be described in the following section.
\subsection{The quantum mechanical configuration space}
The angle variables can be introduced in the quantum system by using the set of functions
\begin{equation}\label{eq-phirep}
\ket{\varphi_1,\varphi_2,\varphi_3} =\!\!\!\!\!\!\! \sum_{n_1,n_2,n_3 \ge 0}\!\!\!\!\!\!\!
e^{i(n_1\varphi_1+n_2\varphi_2+n_3\varphi_3)}\,\ket{n_1,n_2,n_3}\,,
\end{equation}
first introduced in molecular spectroscopy by Sibert and McCoy \cite{Sibe96}. 
These functions are similar to the Bargmann states studied in \cite{Angl01b} in the context of a Bose-Einstein condensate.
This relation is well-known from the context of infinite lattices. There, the sum is taken from $-\infty$ to $\infty$ and corresponds to the representation of Bloch functions in terms of Wannier functions. The angle variables $\varphi_1$, $\varphi_2$ and $\varphi_3$ span the Brillouin zone. However, in this example these functions are not orthogonal due to the fact that for fixed $N$ the sum is finite, 
\begin{multline}
\scal{\varphi'_1,\varphi'_2,\varphi'_3}{\varphi_1,\varphi_2,\varphi_3}\\[2mm]
=\!\!\!\!\!\!\!\sum_{n_1+n_2+n_3=N} \!\!\!\!\!\!\!
e^{-i(n_1(\varphi_1-\varphi'_1)+n_2(\varphi_2-\varphi'_2)+n_3(\varphi_3-\varphi'_3))}\,.
\end{multline}
For a large particle number $N$ the scalar product converges to a delta-comb. There is a considerable deviation for the value of $N=30$, which can play an important role when matrix elements are calculated. But here we  use these functions only for visualization and not for further algebraic manipulations.
The eigenfunctions of (\ref{eq-Hqm}) have the form
\begin{equation}\label{eq-eigexpansion}
\ket{\ef}= \!\!\!\!\!\!\!\sum_{n_1+n_2+n_3=N}\!\!\!\!\!\!\! c_{n_1,n_2,n_3}\,\ket{n_1,n_2,n_3}\,.
\end{equation}
The coefficients $c_{n_1,n_2,n_3}$ can be obtained by a numerical diagonalization in the number basis $\ket{n_1,n_2,n_3}$. The eigenstates in the angle representation (\ref{eq-phirep}), i.e.\ the wave functions, are given by a Fourier series:
\begin{equation}\label{eq-eigf}
\scal{\varphi_1,\varphi_2,\varphi_3}{\ef} = \!\!\!\!\!\!\!\sum_{n_1+n_2+n_3=N}\!\!\!\!\!\!\!
c_{n_1,n_2,n_3}\,e^{i(n_1\varphi_1+n_2\varphi_2+n_3\varphi_3)}\,.
\end{equation}
Finally, we can reduce the number of degrees of freedom in this representation by using the same coordinate transformation as in the classical case of Eq.~(\ref{eq-vartrans}). This leads to the expression
\begin{multline}\label{eq-eigfred}
\ef(\psi_1,\psi_2)=\scal{\psi_1,\psi_2}{\ef}\\
=e^{iN\vartheta}\sum_{n_1+n_3\le N}c_{n_1,N-n_1-n_3,n_3}\,e^{i(n_1\psi_1+n_3\psi_2)}\,.
\end{multline}
The global phase factor $e^{iN\vartheta}$ can be ignored in the following considerations. It 
must be emphasized that the sum includes only a finite number of terms due to the finite 
number of combinations of numbers $n_1$, $n_2$ and $n_3$ which sum to $N$. Therefore the Fourier expansion in Eq.~(\ref{eq-eigfred}) has only a finite resolution. For a very small value 
of $N$, this sum has just a few terms, so that only the coarse grain structure can be explored;
accordingly, the eigenfunctions $\ef(\psi_1,\psi_2)$ show only diffuse structures.
In our example, the configuration space of the reduced system is the two dimensional torus $T^2$ with total volume $4 \pi^2$. The total number of basis states for a
given number of particles $N$ is $L = (N+1)(N+2)/2$. Accordingly, the eigenfunctions
which are linear combinations of the $L$ basis functions can
only show patterns with a resolution of the order $4 \pi^2 /L$ in the area or a resolution of
the order $2 \pi /N$ in each direction. With $N=30$ we have $L=496$ eigenstates,  giving a resolution of approximately $0.07\pi$ in each direction.

The reinterpretation of the expansion of an eigenstate into number states as a
Fourier series on the toroidal configuration space has the following
semiclassical interpretation, where we
write for the moment $\hbar$ explicitly into the equations:
If one naively quantizes the classical canonical variables $(\varphi_k,I_k)$ using the Schr\"odinger quantization  $[\hat \varphi_k,\hat I_l] = i\hbar\delta_{kl}$, which imposes $\hat I_k =(i\hbar)^{-1} \partial/(\partial\varphi_k)$, then functions $f(\varphi_1,\varphi_2,\varphi_3)$ can be interpreted as wave functions in coordinate space. Of course, the Schr\"odinger quantization is correct only in Cartesian coordinates  and it does not commute with canonical transformations, in general yielding errors of the order of $\hbar^2$. Therefore the results have to be interpreted semiclassically. Note that due to our symmetric introduction of the quantum-classical correspondence in Eq.~(\ref{eq-Inrel1}), the errors to first order in $\hbar$ cancel identically. 
Because of these considerations we call the wave function from Eq.~(\ref{eq-eigfred}) the
{\it semiclassical wave function}.

In many semiclassical investigations, Husimi functions are used to relate
quantum wave functions of eigenstates to structures in the classical phase
space. This is the appropriate and natural procedure if the usual position
and momentum coordinates are used. It is less clear and in addition not
necessary in our case where the whole dynamics is treated in action-angle
variables. Let us explain this point in some detail: The description of the
system by a Hamiltonian of the functional structure of 
Eqs.~(\ref{eq-Hqm})--(\ref{eq-Hqmb}) in the 
quantum case or Eq.~(\ref{eq-Hcl}) in the classical case only makes sense for bound systems,
it is not appropriate to describe scattering systems. Therefore we restrict
the following discussion to bound states only. For any bound eigenstate in 
the standard position space, there must be the same amount of wave running in
one direction and in the opposite direction, otherwise it would not be a
bound stationary state. Accordingly, the wave function can be chosen real.
The phases of the wave function do not play any important role and do not help
for the classification of the states. The canonically conjugate 
momenta have continuous values and Wigner or Husimi functions are
defined without any problem on the classical phase space and indicate in
many cases to which structure in the classical phase space some particular 
quantum state belongs.

The situation is very different in action angle variables. Here the 
configuration space is a torus with its very different global topology.
This causes great difficulties to define the usual Wigner or
Husimi functions. Because of the periodicity of the configuration coordinates,
the corresponding canonically conjugate variables (here the actions) only
have discrete values in the quantum dynamics. This makes it very tricky to
convert the wave function into something defined on the continuous classical
phase space. On the other hand, we do not really need to do this, since we have
the following simpler method to squeeze out of the wave functions 
information on the classical actions.
Waves propagating in one direction on a torus always return to the starting point.
Accordingly, wave functions for a bound state can have -- and in fact in most cases do
have -- strong running wave contributions and the phase of the function is
essential and will be analyzed to help in the classification of the state.
In a semiclassical spirit the phase of a wave function can be interpreted as
a classical action integral and accordingly the gradient of the phase function
gives the value of the canonically conjugate momentum which in this case is 
the action. If there is a sufficiently large patch of configuration space 
where the phase function comes close to a plane wave, then its gradient
indicates the value of the actions which is represented by this part of 
the wave function. This provides a kind of lift of the wave function from
configuration space into phase space. If there are closed loops on the torus 
along which the phase function is very regular (and this usually happens
along density crests which run along the classical organizing center as 
will be explained in detail in section~\ref{sec-semiclass}) 
then we interpret this as
representing a motion of almost constant action along this loop. This idea
is used to get longitudinal quantum numbers introduced in section~\ref{sec-semiclass}.
\section{Classical dynamics and coupling schemes}\label{sec-class}
Before we relate individual quantum states to guiding centers of
the classical dynamics, we must get an overview of the classical 
dynamics and its skeleton. As an example, we discuss the classical dynamics 
for $N=30$, i.e.\ for the value
$31.5$ of the classically conserved total action $K$. In the following,
we choose parameter values $\omega_1 = -\omega_3 = 0.1
$, $\omega_2 = 0$, $x_1=x_2=x_3=0.1$ and $k_{1,2}=k_{2,3}=0.5$, which
lead to a quantum mechanical energy interval of $[23.907,
96.393]$. The classical reduced system exists in the energy interval
$[22.476, 99.1]$.
Furthermore, we measure all energies with respect to the quantum
mechanical ground state of $H$ in Eq.~(\ref{eq-Hqm}), i.e.\ we subtract
the quantum mechanical zero point $H_0(1/2,1/2,1/2)=0.075$ from the
classical energies in order to facilitate the comparison between
classical and quantum dynamics. To represent the classical dynamics
graphically, we show Poincar\'e sections in planes $\psi_1=0$ with positive 
orientation $\dot{\psi}_1>0$. 
If an initial condition $(\psi_2, J_2)$ is chosen in the Poincar\'e section,
then we first have to reconstruct the four corresponding coordinates in the phase
space in order to start a trajectory of the flow through this point. The two 
coordinates $\psi_2$ and $J_2$ coincide with the given coordinates in the 
domain of the Poincar\'e map. The coordinate $\psi_1$
is obtained by the intersection condition and the remaining coordinate $J_1$
is calculated by an inversion of the 
Hamiltonian function~(\ref{eq-Hclred}) with respect to the coordinate $J_1$  
for a fixed value of the energy and for the known values of the other three coordinates. 
Here some care is necessary since this inverse function is multivalued.
First we fix one orientation of the domain, i.e. we always search for solutions
with $d \psi_1 / dt > 0$. In principle there can be several solutions with
the same orientation and then it is necessary to ensure that all initial
points used belong to the same branch. 
Poincar\'e sections in planes $\psi_2= \text{constant}$ look very similar to 
the ones in planes $\psi_1=\text{constant}$. Therefore it is sufficient to restrict 
ourselves to sections in $\psi_1=0$ only.

If the whole dynamics were governed by $H_0$, then all actions would
be
constants of motion and all Poincar\'e sections would be foliated
by invariant lines $J_2=\text{constant}$. Including the interaction $W$ between the 
wells (modes) into the dynamics
has the following effects. In regions of the phase space, where none
of the resonances contained in $W$ has an important effect, the dynamics
is in the KAM regime (see the extensive discussion of soft chaos in
chapter
9 of \cite{Gutz90}) and a large fraction of the phase space 
volume is still filled by invariant lines, which are continuous
deformations
of the invariant surfaces $\vec{J}=\text{constant}$ of the unperturbed $H_0$ dynamics.
We call such invariant surfaces primary tori.
This happens mainly in regions of phase space where the effective 
frequencies
\begin{equation}\label{eq-omegaeff}
\omega^\text{eff}_j = \frac{\partial H_0}{\partial I_j} 
\end{equation}
are far from simple rational ratios, for which there is a corresponding
resonance coupling in $W$, as explained in the next paragraph. For our particular
choice of coupling terms in $W$, only 1:1 resonances are relevant.

The effect of the coupling terms between the different modes can be described in the
following way.
Each term contains a cosine function whose argument is a difference
between angles of the original degrees of freedom or one angle of the
reduced system, see Eqs.~(\ref{eq-Hcl}) and (\ref{eq-Hclred}). 
Because in our special case the arguments are differences of two angles
with the same weight, we say that these terms describe 1:1 resonant
interactions between 
the two degrees of
freedom. The right hand sides of the
Hamiltonian equations of motion (\ref{eq-diffhama}) and (\ref{eq-diffhamb})
for the angles $\psi_k$ ($k=1,2$) of the reduced system,
\begin{equation}
\Diff{\psi_k}{t} = \diff{H_0}{J_k} + \diff{W}{J_k}\,,
\end{equation}
contain two contributions. The first consists of the difference of two effective
frequencies from Eq.~(\ref{eq-omegaeff}), and the second is the derivative of the
coupling terms with respect to the action, which 
contains cosine functions. 
First, let us assume that we change some parameter, e.g.\ $k_{1,2}$, to see
how coupling sets in. Further we assume that the difference between the effective 
frequencies, i.e. the angle independent term on the right hand side, is 
different from zero. Let us say it has the value $\nu \ne 0$. For a small 
value of $k_{1,2}$ the angle dependent terms are not able to cancel $\nu$
regardless of the value of the angles. The angle dependent terms have the 
maximal absolute value for angle values $0$ and $\pi$ because of the
dependence on cosine functions. When $k_{1,2}$ increases, then at one point 
it reaches a value, where the angle dependent terms are just able to cancel
$\nu$. Then the angle $\psi_k$ of the reduced system stops, $\psi_k(t) = \text{constant}$,
and we call this {\it frequency locking}. This necessarily happens for 
angle values where the cosine functions
have maximal absolute value, i.e. where the angles are 0 or $\pi$. Whether the
appropriate angle values are 0 or $\pi$ depends on the signs of $\nu$ and of
the terms in front of the cosine functions. When the value of $k_{1,2}$ is
further increased, then there is a whole interval of angle values where
locking is possible. The actual dynamics of the locked motion then performs 
small oscillations
around the angle values 0 or $\pi$. This will be seen in the numerical 
results of the classical dynamics. In the quantum dynamics
the fluctuations around the coupling point of the angles are quantized and
give rise to a discrete set of transversal quantum numbers, see section IV.

If only one of these resonant couplings is strong, then the dynamics is
still close to integrable, and a large part of the phase space
volume is filled by invariant tori, which show up as invariant lines in
the Poincar\'e sections. However, due to the rearrangement of phase
space structures by the resonant coupling, the invariant surfaces in phase space
 are no longer primary tori, i.e.\ are no longer continuous deformations 
of invariant surfaces of the $H_0$ dynamics.
Large bundles of secondary tori appear which are organized around
periodic
orbits (in this case stable, elliptic) representing the {\it guiding
centers} 
for the new nonlinear modes. There are also corresponding
unstable periodic orbits, which in the integrable case are represented
by separatrix
crossings in Poincar\'e sections. In the nonintegrable cases, the
separatrices
break and turn into homoclinic tangles, which become the central
structures of chaotic strips. However, if only one resonant coupling
has a strong effect and the others are not important, then
the chaos strips are very thin and they still appear almost like
separatrices.

If two or more linearly independent resonant couplings are strong,
then chaos on large scales can appear.
These regions in phase space are resonance overlap zones
\cite{Chir79}. However, also in strongly chaotic regions of phase space
there are still simple short periodic orbits (in this case unstable,
normal hyperbolic or inverse hyperbolic) which act as guiding centers
of the flow. Then the dynamics is chaotic but nevertheless 
the flow follows some guiding center on the
average. This average flow
is relevant for the comparison with quantum dynamics.
Thus,  also in the classically chaotic case we may find surprisingly
simple and clean structures in a large part of the quantum wave
functions. In such cases it can be appropriate to imagine
simple idealized classical guiding centers and interpret the quantum 
states as quantum excitations of these idealized structures.

Let us give a short estimate of the size of structures which are relevant
for our semiclassical considerations. The range of action values is limited
between 0 and $K$ due to Eq.~(\ref{eq-Kcl}), the angle can vary over an 
interval of length $2 \pi$.
For each particular plot only a part of this range is energetically 
accessible in reality.
Accordingly the size of the Poincar\'e section is limited by
$2 \pi K$. For semiclassical investigations structures of a size of $\hbar$
or larger are relevant. We always use units in which $\hbar$ has the 
numerical value 1 and also the values of all actions should be interpreted 
as being given in units of $\hbar$. Therefore structures in our Poincare 
plots are of interest in the following, if their size is at least in the order
of one unit of action or has a relative size of $1/K$ compared to the size
of the maximally possible domain of the map.

We perform almost all our calculations for the reduced system. On the 
other hand, the real object of interest is the original system of particles 
in three wells. Therefore we need a fast and easy method to transfer
statements about the reduced system into the corresponding statements about
the original system. We have called this procedure the {\it lift} in the previous
work on molecular systems \cite{Jung04,Jaco99,Jung02}. Let us assume a trajectory of the reduced
system is given and we want to reconstruct the corresponding trajectory of
the original system. The first step of the procedure is the
reconstruction of the cyclic angle. It is done rigorously by using the
Hamiltonian equation of motion
\begin{equation}
\Diff{\vartheta}{t} = \diff{H}{K}\,.
\end{equation}
The right hand side of this equation does not depend on $\vartheta$ but
only on the known values of the other coordinates as function of time.
Accordingly we get $\vartheta(t)$ by a simple integration with respect to time.
The experience with the molecular systems has shown that normally it is
sufficient to approximate $\vartheta(t)$ by $t$ times a constant effective 
frequency. In our case $\vartheta$ is the only fast variable of the whole system
and describes a fast oscillation superimposed on the motion of the whole
system. The initial value $\vartheta(0)$ is rather irrelevant.
In contrast, the variables of the reduced system are slow variables
describing the relative motion between the various degrees of freedom of 
the original system.
The next step of the lift procedure is to undo the canonical transformation 
and to go back to
the coordinates of the original system. In this second step the advantage 
of choosing the new actions equal to some of the old actions becomes evident.
The knowledge of the actions in the reduced system and of the constant value of $K$ gives
immediately the values of the old actions, i.e.\ the values of the particle
numbers in the three wells.
Because of this simple connection between the actions of the reduced system
and the actions of the original system we will switch very freely between
the reduced and the original system in the following considerations.

In our case we have in the interaction part $W$ of the Hamiltonian 1:1 couplings
between the degrees of freedom 1 and 2 and between 2 and 3 respectively.
Indirectly this also implies a 1:1 coupling between the degrees of
freedom 1 and 3. Accordingly, we have the following coupling schemes:
\begin{figure}[h]
\begin{center}
\includegraphics[width=8.5cm]{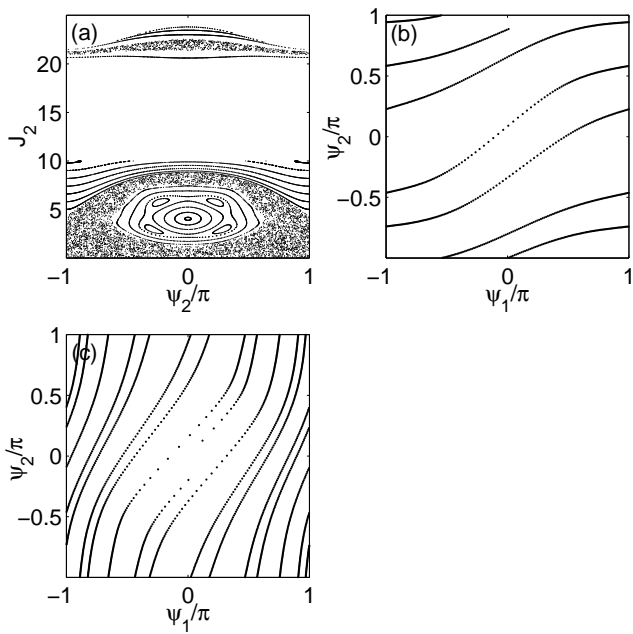}
\end{center}
\caption{\label{fig-class55} The classical reduced system for energy $E=55$. (a) Poincar\'e plot in the plane $\psi_1=0$ for variables $\psi_2$ and $J_2$ with $J_1$ fixed by energy conservation. (b) Trajectory in a primary torus in the lower region of (a) for initial values $(\psi_1,\psi_2,J_2)=(0,\pi,7.5)$. (c) Trajectory in a primary torus in the upper region of (a) for initial values $(0,0,20.6)$. The points of the trajectories are given in equidistant time intervals $\Delta t=0.01$ in order to indicate the velocity by the distance between neighboring points.}
\end{figure}

{\it Type (A)\,}: If the effective frequencies are not very close to each
other, then
no interaction term can cause frequency and phase coupling, and all three
modes run independently with their own effective frequency. This is the
KAM regime with many primary tori, where the motion is of quasiperiodic 
type with three independent frequencies. The organization center of the
reduced 
system is the complete configuration space $T^2$. In Poincar\'e plots, we
see
many invariant lines which are continuous deformations of horizontal
lines
$J_2 = \text{constant}$, i.e.\ of the invariant lines belonging to $H_0$.
This type of motion appears mainly in the middle of the accessible energy
interval for a given particle number. In Fig.~\ref{fig-class55} we give 
some numerical results for the energy $E=55$. Part (a) shows
the Poincar\'e section and parts (b) and (c) show two segments of trajectories
in the reduced configuration space. The domain of the Poincar\'e map in (a) consists of two 
parts. The range of $J_2$ values between approximately 10.2 and 20.5
is not accessible at this energy. At values of $J_2$ around 9, we see many 
primary tori. A segment (five revolutions in direction of $\psi_1$) of a 
typical trajectory belonging to one of them
is shown in part (b) of the figure. In the long run, the trajectory fills 
the whole configuration space quasiperiodically. In these primary tori the
action $J_2$ is smaller than the action $J_1$, so that the trajectories
move faster in $\psi_1$ direction than in $\psi_2$ direction. The opposite
happens on the primary tori lying around $J_2$ values of 21. Here the $J_2$
action is largest and therefore the quasiperiodic trajectories run with
higher speed in the $\psi_2$ direction. (For a numerical example, see a trajectory
segment in Fig.~\ref{fig-class55}(c)).
The other structures seen in Fig.~\ref{fig-class55}(a) belong to other types of motion, 
discussed below.

\begin{figure}[htb]
\begin{center}
\includegraphics[width=8.5cm,  angle=0]{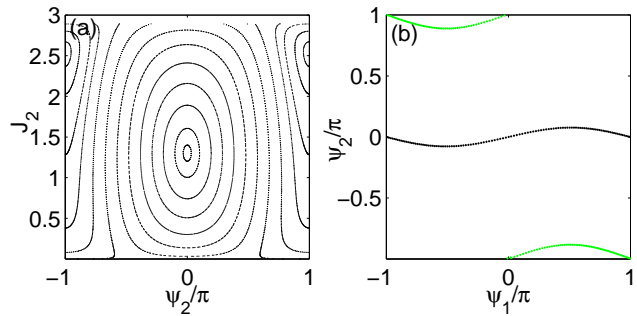}
\end{center}
\caption{\label{fig-class80}(Color online) The classical reduced system for an energy $E=80$. (a) Poincar\'e plot as in Fig.~\ref{fig-class55}(a). (b) Periodic orbits crossing the Poincar\'e section in the centers of the KAM islands (black) for initial values $(\psi_1,\psi_2,J_2) = (0,0,1.3)$ and at the border (green) for initial values $(0,\pi,2.6)$.}
\end{figure}

{\it Type (B)\,}: If the effective frequencies of modes 2 and 3 are
close but that of
the first mode is not close, then we expect that modes 2 and 3 are locked but
mode
1 is independent. The motion is then quasiperiodic with two
independent 
frequencies. The organization center in the reduced system is a one
dimensional
curve with $\psi_2 = \text{constant}$. In
Poincar\'e plots in the plane $\psi_1 = 0$, we see secondary
islands. This motion appears mainly for high energies. Figure~\ref{fig-class80} gives some
numerical results for $E=80$. Part (a) shows a Poincar\'e section, again in 
the plane $\psi_1=0$, and part (b) shows two periodic orbits in the
configuration space. The motion at the upper end of the accessible energy interval is close to
integrable. At a very high energy, motion in $\psi_1$ direction is 
preferred, since the linear frequency $\omega_1$ of original mode 1 is higher than 
the frequency $\omega_3$ of mode 3. For decreasing energy, the KAM island around the center
at $\psi_2 =0$ increases in size while the one around $\psi_2 =\pi$
decreases. The central periodic orbit around $\psi_2=0$ remains stable 
for energies down to approximately $E=45$, while the other one soon becomes 
unstable and its KAM island disappears. In Fig.~\ref{fig-class55}(a) we see clearly
the large KAM island belonging to the organization center $\psi_2=0$,
with center  at $J_2=4$.
In contrast to the idealized organization center $\psi_2=0$, the exact one is a periodic trajectory running in $\psi_1$ showing small wiggles in $\psi_2$ direction around the average value $\psi_2=0$. 
However for our considerations it is simpler and completely satisfactory to replace this
true organization center, the true periodic orbit, by an idealized
organization center, for which we just take the straight line $\psi_2=0$.
The reader might remember the previous discussion of the onset of angle 
coupling and the values of the angles at which coupling sets in. In the spirit
of this previous discussion we define the idealized organization center as
the subset of the configuration space defined by the angle restrictions 
exactly at the onset of the
corresponding coupling scheme. Also the idealized semiclassical wave functions are given
with respect to the corresponding idealized organization center.

\begin{figure}[htb]
\begin{center}
\includegraphics[width=8.5cm,  angle=0]{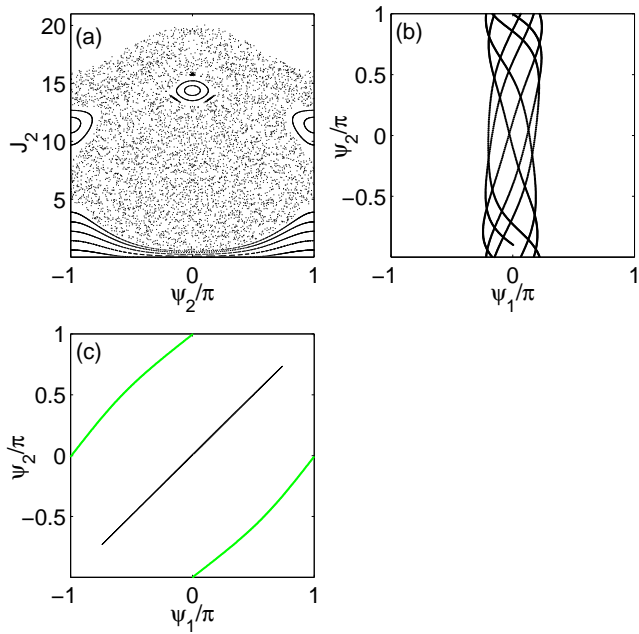}
\end{center}
\caption{\label{fig-class40}(Color online) The classical reduced system for an energy $E=40$. (a) Poincar\'e plot as in Fig.~\ref{fig-class55}(a). (b) Quasiperiodic orbit with initial values $(\psi_1,\psi_2,J_2)=(0,\pi,2)$. (c) Two periodic orbits: one oscillating along
the diagonal with starting point $(0,0,14.4)$ (black) and the other rotating around the line $\psi_1 = \psi_2+\pi$ (green) with starting point $(0,\pi,11.5)$.}
\end{figure}

{\it Type (C)\,}: If the effective frequencies of modes 1 and 2 are
close, but that of
the third mode is not close, then we expect that modes 1 and 2 are locked but
mode
3 is independent. Then the motion is again quasiperiodic with two independent 
frequencies. In the reduced system, the organization center is a one
dimensional
curve which can be idealized by a line $\psi_1 = \text{constant}$, where the 
constant usually is $0$ or $\pi$ according to the discussion in the
beginning of this section. 
The periodic orbit itself running in the $\psi_2$
direction is almost impossible to find in Poincar\'e maps with plane of 
intersection
$\psi_1=0$, since it violates the transversality of the map. However, when it is stable, 
then there is a bundle of invariant tori around it. In Poincar\'e plots
in the planes $\psi_1 = 0$, these invariant tori appear as lines extending
over all values of $\psi_2$. In Fig.~\ref{fig-class55}(a) they are the lines at the highest values
of $J_2$.
In Fig.~\ref{fig-class40}, we show some numerical results at energy $E=40$. Part (a) shows the
Poincar\'e map and parts (b) and (c) show trajectories in configuration space.
In Fig.~\ref{fig-class40}(a) the lines at small values of $J_2$ belong to the tori around the
organization center $\psi_1=0$. Figure~\ref{fig-class40}(b) shows a segment of a typical
quasiperiodic orbit on one of these tori. While running monotonously in the
negative $\psi_2$ direction, it oscillates in $\psi_1$ around the value 0.

{\it Type (D)\,}: If the effective frequencies of modes 1 and 3 are very
close, then
also the weak indirect tunneling processes between modes 1 and 3 can cause
coupling.
If the frequency of mode 2 is far from this common frequency, 
then mode 2 runs independently. The corresponding organization
center
in the reduced system is the line $\psi_1 = \psi_2 + \text{constant}$, where
again 
this constant is usually 0 or $\pi$. In Poincar\'e plots in planes 
$\psi_1 = 0$, we see secondary islands.
In Fig.~\ref{fig-class40}(a), the two KAM island of moderate size with centers at $\psi_2=0$,
$J_2=14$ and $\psi_2 = \pi$ and $J_2=11.5$ respectively represent this type
of motion. The two periodic orbits belonging to the centers of these two
KAM islands are shown in Fig.~\ref{fig-class40}(c). One oscillates along the diagonal and 
the other rotates around along the line $\psi_1 = \psi_2 + \pi$.

\begin{figure}[htb]
\begin{center}
\includegraphics[width=8.5cm,  angle=0]{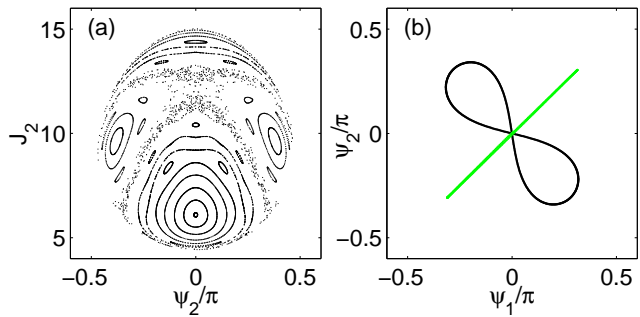}
\end{center}
\caption{\label{fig-class27}(Color online) Classical reduced system for energy $E=27$. (a) Poincar\'e plot as in Fig.~\ref{fig-class55}(a). (b) Trajectories through the stable fixed point (black, double loop) with initial values $(\psi_1,\psi_2,J_2)=(0,0,6.1)$, and the unstable fixed point (green, along the line $\psi_1=\psi_2$) with initial values $(0,0,12.1)$.}
\end{figure}

{\it Type (E)\,}: If all three effective frequencies are close, then
there are two
possibilities: 

(E1): There is coupling between all three modes and the
idealized organization center in the configuration space of the reduced
system is a (fixed)point. The actual trajectories oscillate around this
coupling
point and the relative angles $\psi_k$ do not rotate around the whole configuration
torus. This behavior, which dominates at very small energy, is shown
in Fig.~\ref{fig-class27} at energy $E=27$. Only a limited range of $\psi_2$ values around
the point zero is energetically accessible. The same also holds for $\psi_1$.
Rotations around the configuration torus in
either direction or the diagonal become possible only for a higher energy. One of the organizing centers 
is represented in the Poincar\'e plot by a stable fixed point which lies at the center of 
the large KAM island
shown in Fig.~\ref{fig-class27}(a), and which is shown in configuration space in 
Fig.~\ref{fig-class27}(b) as the figure-of-eight orbit mainly oscillating in the antidiagonal direction. 
The other organizing center is an unstable periodic orbit belonging to the
unstable fixed point near $\psi_2=0$, $J_2=12$ in the Poincar\'e plot.
In the configuration space plot of Fig.~\ref{fig-class27}(b), it is the orbit oscillating in
the diagonal direction. At this energy, all trajectories in configuration
space oscillate around the point $(0,0)$, which acts as point organizing
center. The two periodic orbits of Fig.~\ref{fig-class27}(b) then act as guiding structures for
these fluctuations around the organizing center. Topologically speaking, all trajectories are contractible to a point on the configuration torus at very low energy. 
At the lower end of the accessible energy interval, the dynamics starts as almost integrable 
and for this case the invariant manifolds of the unstable fixed point mentioned above lie close 
to a figure-of-eight shape separatrix
in the Poincare section. For increasing energy the system moves further 
away from integrable and the separatrix breaks and turns into a homoclinic
tangle which is the central structure of a chaos strip. In Fig.~\ref{fig-class27}(a) for energy 
$E=27$ this chaotic layer still has moderate size. For higher energy it 
grows rapidly and turns into the large chaotic sea seen in Fig.~\ref{fig-class40}(a) at 
energy $E=40$.

(E2): The couplings break and reestablish intermittently, and the
dynamics shows large-scale chaos. The appearance of chaos in the case of two
independent resonant interactions becoming active is a demonstration of
Chirikov's point of view of chaos being caused by resonance overlap
\cite{Chir79}.
In Poincar\'e plots, we see large-scale chaos and eventually embedded in it
remnants of islands and regular structures.
The beginning of chaos for small energies can be seen in Fig.~\ref{fig-class27}(a); chaos on
a large scale is evident in Figs.~\ref{fig-class55}(a) and \ref{fig-class40}(a).
\section{Classification of semiclassical wave functions}\label{sec-semiclass}
In this section, we show examples of wave functions belonging to the
various classes of motion described in the previous section. 
Our method of classification has been developed in \cite{Jung04,Jaco99,Jung02,Sibe96} especially
for Hamiltonians given quantum mechanically in raising and lowering operators
and classically in action-angle variables. An analysis in a similar spirit 
of wave functions in the usual position space is rather common in molecular
physics, two representative examples are \cite{Jost99, Azza03}. In contrast to the procedure in action-angle space, the procedure in regular 
position space can be extended to scattering resonances, see \cite{Gome89}.

Our strategy of
classification is as follows: First we expand the eigenstates $\ket{\ef}$ in the representation 
$\ket{\psi_1,\psi_2}$, according to Eq.~(\ref{eq-eigfred}). This representation of the eigenstates in the reduced configuration space is in complete analogy to the classical configuration space spanned by the angle variables $\psi_1$ and $\psi_2$ and therefore allows  a direct comparison between the classical and quantum system. We refer to these eigenfunctions  $\ef(\psi_1,\psi_2)$ as the {\it semiclassical wave functions} in order to indicate this resemblance. We then check whether 
the density of the semiclassical wave function in the reduced configuration space
resembles the structure of one of the organization centers described in the
previous section (types (A) -- (E2)). I.e.\ we check, whether the density is distributed
over the whole configuration space without clear nodal structures (type (A)), is concentrated along a few lines in the $\psi_1$ direction (type (B)),
in the $\psi_2$ direction (type (C)) or in the diagonal direction (type (D)), is 
organized around the point center $(0,0)$ (type (E1)) or shows random interferences between the pattern of different organization centers leading to irregular structures (type (E2)). 

We call states, for which the density is located in a single crest along the organizing center, a 
transverse ground state to this organization center.
In transversely excited states, the density is concentrated along various
copies of the organization center, where these various copies are 
displaced relatively to each other and the wave function shows
nodal structures between them.  
In addition, we look for the phase advance in
directions in or parallel to the organization structure. The phase 
function must be continuous along curves which do not cross nodal lines.
Recall that the phase function can have singularities only in zero points 
of the density. Accordingly, the curve along a crest of high density
must be a curve of continuous phase. Then the
phase advance of such a curve
must be some integer multiple of $2 \pi$, say $\mu_l \cdot 2 \pi$, and this
number $\mu_l$ serves as one quantum number of the state. These longitudinal
quantum numbers, together with the transverse quantum
numbers given by the nodal structures, provide a complete
set of quantum numbers characterizing the state relative to its
organization center. We expect all states which can be related to an 
organization center to be close to a product of a plane wave in
the longitudinal
direction of this organization center and an oscillator function in
transverse directions.

In states belonging to classically chaotic motion,
we do not see a simple and clear pattern in the density nor in the phase.
Accordingly we are not able to give any assignment by quantum numbers
to such states.

In the following, the eigenstates $\ef_k$, $k = 1,\,2,\,\dots 496$ are sorted by increasing energy starting with the label 1 for the eigenstate with lowest energy. 
\subsection*{Point organization center (Type (E1))}
\begin{figure}[htb]
\begin{center}
\includegraphics[width=8cm]{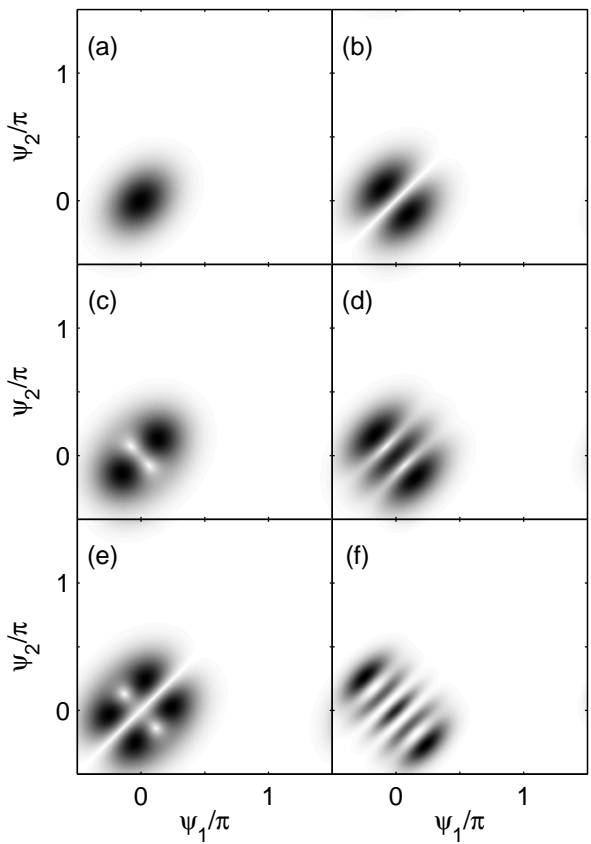}
\end{center}
\caption{\label{fig-punktzentrum} Gray scale plot of the squared modulus of the (a) ground state 
$\ef_1(\psi_1,\psi_2)$ and the excited states (b) $\ef_2$, (c) $\ef_3$, (d) $\ef_4$, (e) $\ef_5$ and (f) $\ef_9$. White color corresponds to low density and black to the highest density. The range of the $\psi_k$ is $[-\pi/2, 3\pi/2]$.}
\end{figure}

We start our analysis of the semiclassical wave functions at the lower end of the accessible energy interval. Since the Hamiltonian is dominated by quadratic anharmonicities, the smallest energy is realized by distributing the total excitation of 30 quanta (particles) evenly over the 3 basis modes (potential wells). In the classical picture, this corresponds to the case where all three actions $I_k$ are close to each other. Thus the three effective frequencies (\ref{eq-omegaeff}) are very similar and frequency and phase locking is easily established by the resonant coupling terms in the Hamiltonian as explained in
the beginning of section~\ref{sec-class}. In the classical configuration space, this mechanism restricts the trajectories to a small region of configuration space (cf. Fig.~\ref{fig-class27}).
This behavior is confirmed in the quantum case. Here, the wave functions are organized around a point, as can be seen in Fig.~\ref{fig-punktzentrum}, where the ground state and various excited states are plotted. State $\ef_1$ is the ground state in this class. In this case the ground state of an organization center coincides with the energetic groundstate $\ef_1$ of the whole system, but we will assign also a groundstate for the other types of guiding centers.
The state $\ef_2$ is the first transversal excitation in the antidiagonal direction, while the state $\ef_3$ represents the first transversal excitation in the diagonal direction. The
state $\ef_4$ represents the second transversal excitation in the antidiagonal direction, and the state $\ef_5$ is the combination of one transverse excitation in the diagonal and
one in the antidiagonal direction. State $\ef_9$ is the fourth excitation in the
antidiagonal direction. A point center does not have any longitudinal
directions. Accordingly, there are no phase advances in longitudinal
directions to be counted for the assignment and any state of this class 
is characterized by the two transverse
excitation numbers $(\mu_{td}, \mu_{ta})$, one in the diagonal direction and one in the antidiagonal direction.
Thus we show only the density plots without the phases in Fig.~\ref{fig-punktzentrum}. 

In this scheme, the six states $\ef_1$, $\ef_2$, $\ef_3$, $\ef_4$, $\ef_5$ and $\ef_9$ have quantum numbers $(0,0)$, $(0,1)$, $(1,0)$,
$(0,2)$, $(1,1)$ and $(0,4)$, respectively.
Note that the direction of excitation corresponds to the direction of 
oscillation of the classical periodic orbits shown in Fig.~\ref{fig-class27}(b).
The classical periods of these two orbits are $T_a= 6.112$ for the antidiagonal
one and $T_d=3.502$ for the diagonal one. The quantum excitations in the
corresponding direction increase the energy of the state by the classical
frequency $\omega = 2 \pi/T$, where $T$ is the period of the orbit taken
at an intermediate energy.

The quantum-classical correspondence can be described in the following way: All three original modes are frequency locked and the phases fluctuate around the coupling point.
The motion is similar to the one in a two dimensional anharmonic oscillator
centered around the point $(0,0)$. This oscillator has its own normal modes
and the states presented in Fig.~\ref{fig-punktzentrum} can be interpreted as some of the low 
lying excitation of this oscillator and described by the excitation numbers 
of these normal modes. However, the reader should not confuse these modes
of fluctuations around coupling points with the modes which are used to 
formulate the original Hamiltonian in Eqs.~(\ref{eq-Hqm})--(\ref{eq-Hqmb}). Compare also with the 
discussion of the onset of coupling given in section~\ref{sec-class}.

The wave functions in this class are therefore close to two dimensional oscillator functions and can be described approximately by
\begin{equation}
\ef_{\mu_{td},\mu_{ta}}(\psi_1,\psi_2) \approx e^{iN\vartheta}\,\chi_{\mu_{td}}(\psi_1+\psi_2)\;
\chi_{\mu_{ta}}(\psi_1-\psi_2)\,,
\end{equation}
where the functions $\chi_n(x)$ are eigenfunctions of a one dimensional oscillator with  harmonic and anharmonic contributions.
It is interesting to see what this means in the original coordinates $\varphi_k$, $I_k$. Using the transformation (\ref{eq-vartrans}), one obtains for the idealized eigenfunctions
\begin{multline}
\ef_{\mu_{td},\mu_{ta}}(\varphi_1,\varphi_2,\varphi_3) \\\approx e^{iN\varphi_2}\, 
\chi_{\mu_{td}}(\varphi_1+\varphi_3-2\varphi_2)\;
\chi_{\mu_{ta}}(\varphi_1-\varphi_3)\,.
\end{multline}
All three degrees of freedom are entangled for this type of guiding center. The entanglement is the quantum analog of the phase locking in the classical picture.
Altogether, we can assign 29 of the 496 eigenstates to this class of functions.
\subsection*{Organization center $\bm{\psi_1 = 0}$ (Type (C))}
\begin{figure}[htb]
\begin{center}
\includegraphics[width=8cm,  angle=0]{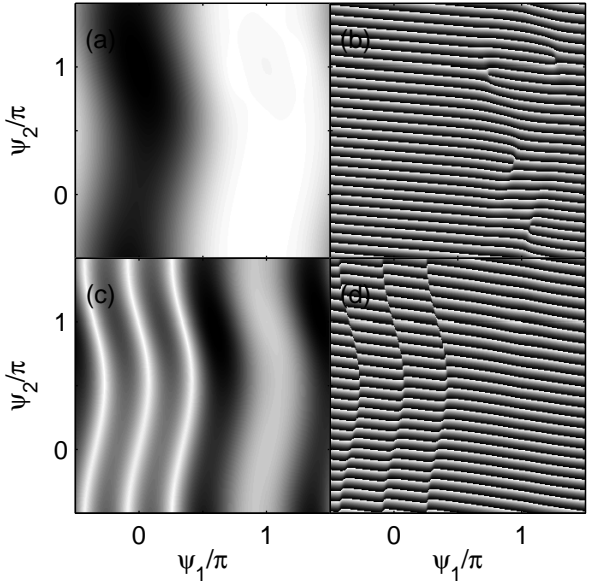}
\end{center}
\caption{\label{fig-psi10} Plot of the eigenfunctions $\ef_{461}$ and $\ef_{433}$ of the 
quantum system belonging to the $\psi_1=0$ guiding center. Plot (a) shows $|\ef_{461}|^2$, 
(b) shows $\arg(\ef_{461})\mod2\pi$, (c) shows $|\ef_{433}|^2$, (d) $\arg(\ef_{433})\mod2\pi$. In the phase
 plots, the degree of darkness from white to black indicates the phase advance from $0$ to $2\pi$.}
\end{figure}

The highest energies for a given number of particles are achieved by putting almost all
excitation into one mode, with the other two modes having very low
excitation. Classically, these two modes have similar effective frequencies, (see Eq.~(\ref{eq-omegaeff})), and therefore they are locked easily.  In Fig.~\ref{fig-psi10} we show 
as examples the densities and phases for the states $\ef_{461}$ and $\ef_{433}$. 
In part (a) we see the density concentrated along the
line $\psi_1 = 0$; thus, the transverse excitation number is $\mu_t =
0$. Along this line, the phase function is almost like a plane wave. The
total phase advance along one cycle around the organization center is $26 \cdot 2
\pi$. Accordingly, the longitudinal excitation number is $\mu_l = 26$. Note that
the phase function has singular points far away from the places of high
density. In part (c) of the figure, we see the density concentrated along four lines 
in the $\psi_2$ direction. The
four density crests are separated by 3 nodal lines, 
which can be seen
very clearly as lines of discontinuities in the phase plot in part (d). Accordingly, the transverse excitation number of state $\ef_{433}$ is 
$\mu_t = 3$. Along the density crests we count the total phase
advance to obtain the longitudinal quantum number $\mu_l = 24$.
The energy distance between two states which differ by one unit in $\mu_l$
and that have the same transverse quantum number is given by the frequency
of the classical organizing center, the periodic orbit (central fiber
in the quasiperiodic motion shown in Fig.~\ref{fig-class40}(b)), taken at an
intermediate energy.

The classical motion behind this class of states is as follows: Mode 3 
runs with its own effective frequency independently of the other modes.
The quantum number $\mu_l$ is its action due to the semiclassical assignment of the phase function $\eta(\psi_1,\psi_2)$ to the classical action integral,
\begin{equation}
\eta(\psi_1,\psi_2) = \int\limits_\gamma \vec{J}\cdot d\vec{\psi}\,,
\end{equation}
with $\vec{J} = (J_1,J_2)$ and $\vec{\psi} = (\psi_1,\psi_2)$.
The phase $\eta(\psi_1,\psi_2)$ is defined by $\ef =|\ef|\exp\{i\eta\}$ and $\gamma$ is the classical guiding center. The path $\gamma$ is simply the line $\psi_1=0$ for this class and the the number of particles in mode 3 can be directly assigned to the quantum number $\mu_l$.
Modes 1 and 2 run locked with
total excitation, i.~e.\ number of particles, $N-\mu_l$, and the quantum number $\mu_t$ characterizes the fluctuations of the coupled motion around the coupling point. 

The eigenfunctions in this class can therefore be written approximately as
\begin{equation}
\ef_{\mu_l,\mu_t}(\psi_1,\psi_2) \approx e^{iN\vartheta}\,e^{i\mu_l\psi_2}\,\chi_{\mu_t}(\psi_1)\,.
\end{equation}
Now we transform this expression back to the original coordinates, where we can interpret the actions $I_k$ directly as the number of particles in the potential well $k$. Using again transformation (\ref{eq-vartrans}), we can write the idealized wave functions of this class as
\begin{multline}
\ef_{\mu_l,\mu_t}(\varphi_1,\varphi_2,\varphi_3) \\\approx e^{i\mu_l\varphi_3}\,e^{i(N-\mu_l)\varphi_2}
\,\chi_{\mu_t}(\varphi_1-\varphi_2)\,.
\end{multline}
This type of wave function shows entanglement between modes 1 and 2, while mode 3 separates. The number of particles in mode 3 is given by $\mu_l$ while the transversal quantum number $\mu_t$ describes the transversal excitation of the organization center. 
A total of 51 eigenstates can be assigned to this class of functions.
\subsection*{Organization center $\bm{\psi_2 = 0}$ (Type (B))}
The states of this class look very similar to those in the previous subsection, 
only with the roles of the modes
1 and 3 interchanged and hence with $\psi_1$ and $\psi_2$ interchanged. 
However, there is no perfect symmetry between classes C and B because
there is no perfect equality between the modes 1 and 3. Remember that
$\omega_1 = -\omega_3 \ne \omega_3$. This small perturbation of the symmetry 
is responsible that the states of class B loose their characteristics under
smaller transverse excitations as the ones for class C. Accordingly we can 
assign less states to class B, namely 42 only, than we have assigned to 
class C.

\subsection*{Organization center $\bm{\psi_1 = \psi_2}$ (Type (D))}
\begin{figure}[h]
\begin{center}
\includegraphics[width=8cm,  angle=0]{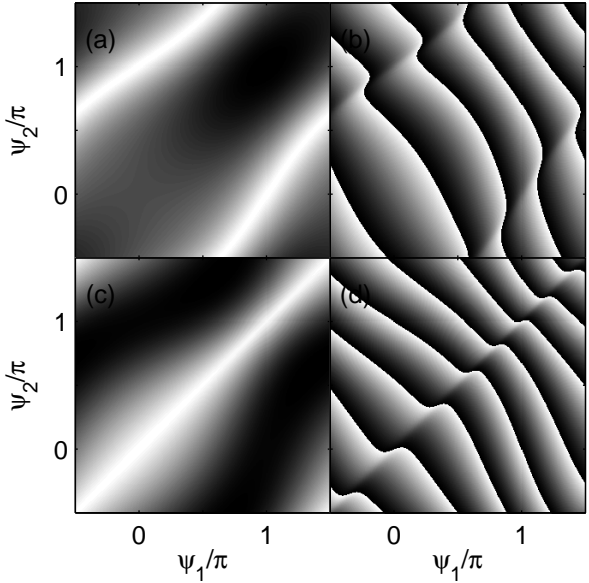}
\end{center}
\caption{\label{fig-psi1psi2} Plot of the eigenfunctions $\ef_{420}$ and $\ef_{359}$ of the 
quantum system belonging to the $\psi_1=\psi_2$ guiding center. Plot (a) shows $|\ef_{420}|^2$, 
(b) $\arg(\ef_{420})\mod2\pi$, (c) shows $|\ef_{359}|^2$, (d) $\arg(\ef_{359})\mod2\pi$.}
\end{figure}

%
If almost all the action $K$ is in mode 2, then modes 1 and 3 have low
actions
and similar effective frequencies, whereas mode 2 has a quite different
effective frequency. Even though the Hamiltonian does not contain a
direct coupling between modes 1 and 3, sometimes the small indirect
coupling is sufficient to cause locking between modes 1 and 3.
Fig.~\ref{fig-psi1psi2} shows the states $\ef_{420}$ and $\ef_{359}$ as two examples of semiclassical wave functions in this class.
The organization center is the diagonal $\psi_1 = \psi_2$. State $\ef_{420}$ has
the transverse quantum number $\mu_t = 0$ relative to this center and
state $\ef_{359}$
has $\mu_t = 1$. The phase functions show that $\mu_l = 6$ for state $\ef_{420}$ and
$\mu_l=8$ for state $\ef_{420}$.
The energy distance between two states, which differ by one unit in $\mu_l$
and that have the same transverse quantum number, is given by the frequency
of the classical organizing center, namely the periodic orbit shown in 
Fig.~\ref{fig-class40}(c) taken at an intermediate energy.

The classical motion carrying these states is the following: The coupled
motion of modes 1 and 3 has the number of particles $\mu_l$ while the rest of the total
excitation $N-\mu_l$ is in mode 2. The transverse
quantum number $\mu_t$ again characterizes the fluctuations around the
coupling point.
For the idealized wave functions of the reduced system, we obtain
\begin{equation}
\ef_{\mu_l,\mu_t}(\psi_1,\psi_2) \approx e^{iN\vartheta}\,e^{i\mu_l(\psi_1+\psi_2)/2}\,\chi_{\mu_t}(\psi_1-\psi_2)\,.
\end{equation}
In the original coordinates, the wave function has the form
\begin{multline}
\ef_{\mu_l,\mu_t}(\varphi_1,\varphi_2,\varphi_3) \\\approx e^{i(N-\mu_l)\varphi_2}\,e^{i\mu_l(\varphi_1+\varphi_3)/2}\,\chi_{\mu_t}(\varphi_1-\varphi_3)\,.
\end{multline}
In these coordinates, mode 2 separates from the other modes which are entangled. The number of particles in mode 2 is given by $N-\mu_l$, while the rest of the particles is in the entangled state of the other two modes, for which the quantum number $\mu_t$ is a measure of the fluctuations around the organization center. We can assign 8 eigenstates to this class of functions.

\subsection*{Organization center $\bm{T^2}$ (Type (A))}
\begin{figure}[h]
\begin{center}
\includegraphics[width=8cm,  angle=0]{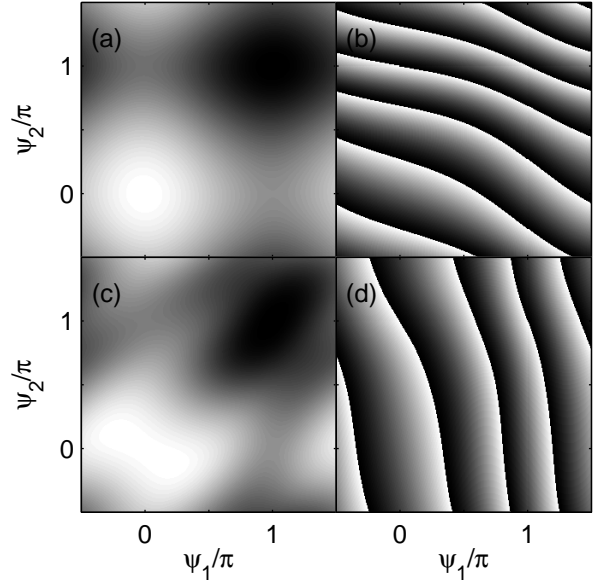}
\end{center}
\caption{\label{fig-torus} Plot of the eigenfunctions $\ef_{401}$ and $\ef_{442}$ of the 
quantum system belonging to the $T^2$ guiding center. Plot (a) shows $|\ef_{401}|^2$, 
(b) $\arg(\ef_{401})\mod2\pi$, (c) $|\ef_{442}|^2$, (d) $\arg(\ef_{442})\mod2\pi$.}
\end{figure}

Fig.~\ref{fig-torus} shows the wave functions of the states $\ef_{401}$ and $\ef_{442}$, which do not show any coupling.  These states belong to
normal mode motion in the original modes. This does not necessarily
mean that they have a constant density, but the density is without any clear structure
and the phase function is close to a plane wave globally.
As the two quantum numbers we count the phase
advances around the two fundamental cycles of the toroidal
configuration
space. In  part (b) of the figure we assign the quantum numbers $\mu_{l1}=2$, $\mu_{l2} =5$ and
from part (d) we read off $\mu_{l1} =4$ and $\mu_{l2}=1$.

These states are described by the classical motion in the following way:
The original
mode 1 has the number of particles $\mu_{l1}$ and original mode 3 has $\mu_{l2}$ particles. The
rest of the excitation $N-\mu_{l1}-\mu_{l2}$ is in mode 2. All
three modes run independently with their own effective frequency. Thus
phase functions of states of this class come close to a basis function (i.e.\ they resemble a plane wave), even though the wave function can be a strong mixture of several basis functions.
The functional form of such states is therefore approximately given by
\begin{equation}
\ef_{\mu_{l1},\mu_{l2}}(\psi_1,\psi_2) \approx e^{iN\vartheta}\,e^{i(\mu_{l1}\psi_1+\mu_{l2}\psi_2)}\,,
\end{equation}
or written in the original coordinates as
\begin{equation}
\ef_{\mu_{l1},\mu_{l2}}(\varphi_1,\varphi_2,\varphi_3) \approx
e^{i\mu_{l1}\varphi_1}\,e^{i(N-\mu_{l1}-\mu_{l2})\varphi_2 }\,e^{i\mu_{l2}\varphi_3}\,.
\end{equation}
These idealized functions factorize and the three degrees of freedom are completely disentangled.
There are 50 eigenstates in this class.

\subsection*{States based on chaotic motion (Type (E2))}
\begin{figure}[h]
\begin{center}
\includegraphics[width=8cm,  angle=0]{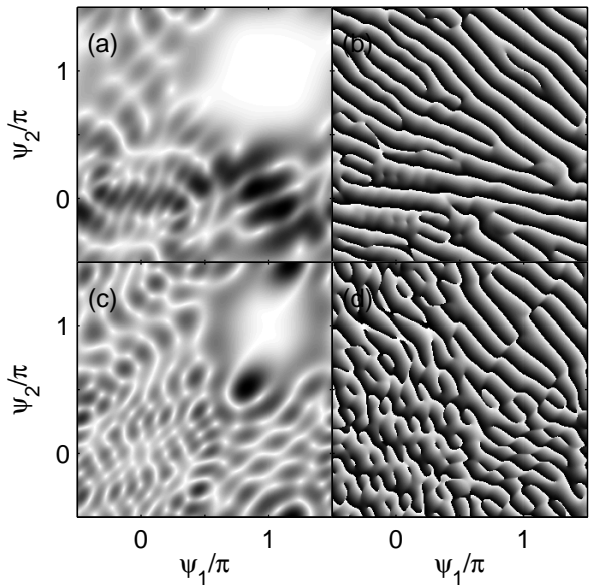}
\end{center}
\caption{\label{fig-chaos} Plot of the eigenfunctions $\ef_{100}$ and $\ef_{146}$ of the 
quantum system belonging to class (E2). Plot (a) shows $|\ef_{100}|^2$, 
(b) $\arg(\ef_{100})\mod2\pi$, (c) shows $|\ef_{146}|^2$, (d) $\arg(\ef_{146})\mod2\pi$.}
\end{figure}

Finally, we give two examples of wave functions where we could not make
any
assignment to one of the organizing centers listed in the previous
section.
Fig.~\ref{fig-chaos} shows the densities and phases  of states $\ef_{100}$ and $\ef_{146}$. Neither in the density plots nor in the phase plots, can we
discover any clean pattern related to one of the organizing centers.
The connection to the classical motion we interpret as follows: In
classical
chaos, any typical trajectory jumps around irregularly between the 
neighborhoods of
various simple periodic orbits and therefore between various types of
motion.
The corresponding quantum wave function should be random interference
patterns of the structures belonging to the various organizing centers
involved in the classical chaotic motion. Sometimes we can demix these
interference patterns by forming appropriate linear combinations of 
several eigenfunctions of the Hamiltonian.

Concluding this section, we are able to characterize 180 of the 496 eigenstates within the scheme of guiding centers given by the classical motion (excluding chaotic motion). Our aim is not a complete assignment of all states, but rather to give an easy visual criterion in order to select states with different types of e.g. entanglement and localization properties as described in this section for each class. For these states, one can use the classical picture in order to understand the quantum mechanical structure, which allows a very intuitive treatment of the states. The above graphical classification of the semiclassical wave functions is not strict and some functions allow ambiguous assignments. Such functions show characteristics of different classes and it is only a matter of degree in which class to put them. For example, the phase functions in Fig.~\ref{fig-psi1psi2} could be interpreted as continuous deformations of plane waves and therefore they could be assigned to type (A) as well.

\section{Comparison of the time dynamics}
Finally, we wish to discuss the implications of our analysis for the time evolution in the 
classical description of the system. The classical system can be interpreted as an array of three  Bose-Einstein condensates where the condensate in each well is described by the Gross-Pitaevskii equation and  where the condensates interact weekly through Josephson tunneling \cite{Smer97, Milb97}.

In the previous section, we have used the classical system only  to provide a tool for the classification of the quantum wave functions, and we have shown how close the quantum eigenfunctions resemble the classical guiding centers. In this section, we look in the other direction. Starting from the classical system, i.e.\ the mean-field equations, we want to ask what information the structure of the quantum system can provide in order to solve the mean-field equations: the analysis of a system of coupled nonlinear differential equations is very involved, while in the quantum system we only have to diagonalize the Hamiltonian numerically and plot the eigenfunctions in configuration space.

Since it is more convenient in this context to speak about complex occupation amplitudes, we introduce the new variables
\begin{equation}
c_k = \sqrt{I_k}\,e^{i\varphi_k}\,.
\end{equation}
In these variables, the classical Hamiltonian (\ref{eq-Hcl}) can be written as
\begin{align}
H &= \sum_{k=1}^3\bigl(\,\omega_k|c_k|^2+x_k|c_k|^4\,\bigr)\nonumber\\&\quad
-\frac{k_{12}}{2}(c_1c_2^*+c_2c_1^*)
-\frac{k_{23}}{2}(c_2c_3^*+c_3c_2^*)\,,
\end{align}
with canonically conjugate variables $(c_k,ic_k^*)$ and corresponding equations of motion
\begin{equation}\label{eq-GPE}
\dot{c}_k= \frac{\partial H}{\partial(ic_k^*)} \quad\Longleftrightarrow\quad
i\,\dot{c}_k^*= \frac{\partial H}{\partial c_k}\;.
\end{equation}
This system of three ordinary differential equations for the complex coefficients $c_k$ is equivalent to the six equations for the angles $\varphi_k$ and the actions $I_k$ with $k=1,2,3$.
The equations can also be derived from the Gross-Pitaevskii equation in coordinate space 
using an expansion of the condensate wave function in Wannier functions \cite{Trom01}. One can also use the eigenstates of the one-particle Hamiltonian, the so-called Wannier-Stark functions, resulting in the disappearance of the linear tunneling terms in the Hamiltonian (\ref{eq-Hqm}), while higher order coupling terms become important \cite{Thom03}.

Now we choose the initial conditions $c_k(t=0)$ by using the 
semiclassical correspondence (\ref{eq-Inrel1}) between the classical actions $I_k$ and the quantum numbers $n_k$ 
of a number state $\ket{n_1,n_2,n_3}$:
\begin{equation}\label{eq-Inrel}
I_k \longleftrightarrow n_k+\frac{1}{2}\,.
\end{equation}
In this way, we can construct initial conditions $c_k(0)=\sqrt{I_k}$, where the action $I_k$ can be interpreted quantum mechanically via Eq.~(\ref{eq-Inrel}) as the number of particles in mode $k$. Furthermore, we can use this correspondence in order to construct initial conditions resembling the properties of the eigenstates of the system. Before we explain this in more detail we first discuss the case of the basis vectors.
\subsection{Basis vectors}
Here we investigate to which extend we can attribute the same characteristics to the quantum mechanical number states $\ket{n_1,n_2,n_3}$ and their classical analog defined by Eq.~(\ref{eq-Inrel}). Accordingly, we define the initial conditions for the time evolution of Eq.~(\ref{eq-GPE}) as
\begin{align}\label{eq-basisb}
&\vec{b}(t=0;n_1,n_2,n_3)=(c_1(0),c_2(0),c_3(0)) =\nonumber\\&\quad= (\sqrt{n_1+1/2},\sqrt{n_2+1/2},\sqrt{n_3+1/2})\,.
\end{align}
In the following we will not explicitly write down the dependence of $\vec{b}(t;n_1,n_2,n_3)$ on the initial condition through the parameters $(n_1,n_2,n_3)$  and simply use $b(t)$.
We fix the three initial phases to zero, which corresponds to zero imaginary part of the $c_k(0)$.
With this initial conditions the time evolution can be calculated numerically, as shown in 
\begin{figure}[htb]
\begin{center}
\includegraphics[width=8cm,  angle=0]{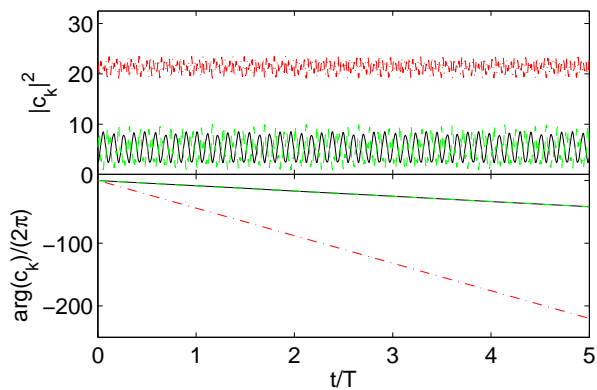}
\end{center}
\caption{\label{fig-GPevolution}(Color online) Time evolution of Eq.~(\ref{eq-GPE}) for an initial condition $\vec{b}(0) = (\sqrt{2.5},\sqrt{5.5},\sqrt{23.5})$. Shown are squared modulus (top) and the phase of the first (solid, black), second (dashed, green) and third (dash-dotted, red) mode. In the phase plot the first and second phase almost coincide and lie above the third phase which has a bigger phase velocity. The time is measured with respect to $T=2\pi/\omega$.}
\end{figure}
Fig.~\ref{fig-GPevolution} for initial values $(\sqrt{2.5},\sqrt{5.5},\sqrt{23.5})$ using Eq.~(\ref{eq-GPE}). In this example the phases of $c_1$ and $c_2$ are locked, while $c_3$ evolves independently. The difference in the amplitudes  between mode 3 and the other two prohibits a coupling. The amplitudes show a quite regular oscillation in all three modes. This is motion of type (C) introduced in section~\ref{sec-class}.
Physically interpreted, the wells 1 and 2 couple through Josephson tunneling and the population between the two wells is exchanged periodically. In contrast, the number of particles of well 3 stays approximately constant and much higher than the population of the other wells. This behavior reflects the well-known macroscopic self-trapping found in the double well potential \cite{Smer97}. Another type of this self-trapping effect in the type (C) dynamics can occur, when wells 1 and 2 have approximately the same population $N/2$ and well 3 is nearly empty.
One can also observe the other types of dynamics in the vectors $\vec{b}(t)$, except type (D), due to the very weak indirect coupling between modes 1 and 3. 
The different time evolutions $\vec{b}(t)$ can be easily assigned to the different guiding centers by looking at the phases: 
\begin{figure}[htb]
\begin{center}
\includegraphics[width=8cm,  angle=0]{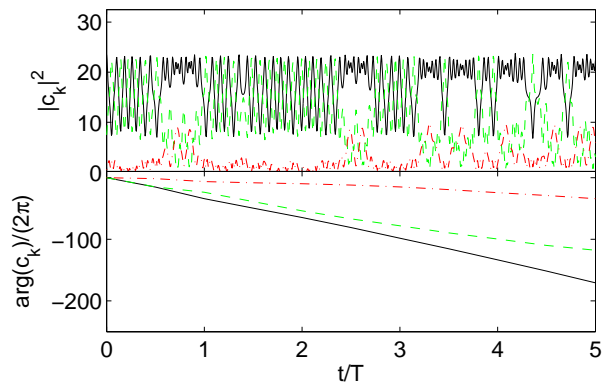}
\end{center}
\caption{\label{fig-GPevolution2}(Color online) Time evolution for an initial condition $\vec{b}(t=0) = (\sqrt{23.5},\sqrt{7.5},\sqrt{0.5})$. Shown are squared modulus (top) and the phase of the first (solid, black), second (dashed, green) and third (dash-dotted, red) mode.  The time is measured with respect to $T=2\pi/\omega$.}
\end{figure}

{\it Type (A)}: All three phases behave independently and the amplitudes oscillate regularly. The individual condensates in the different wells are completely decoupled and the population in each well stays approximately constant.

{\it Type (B)}: The dynamics shows the same behavior as for type (C), but with phase locking between mode 2 and 3.

{\it Type (D)}: This type of motion is difficult to identify, because the indirect phase locking between modes 1 and 3 is very weak. This leads to the effect that the phase velocities of these two phases are very close, but still distinguishable. This is of course not a strict statement, and it depends on how long the time propagation is considered. 
The problems with the classification of this type can also be seen in the quantum case in Fig.~\ref{fig-psi1psi2}. In parts (b) resp.\ (d), the phase singularities are not sharp but rather smooth, so these states  could be assigned to type (A) as well. 

{\it Type (E1)}: In this case all three phases evolve with the same velocity and the amplitudes show similar regular oscillations as in types (B) and (C) for two locked phases.

{\it Type (E2)}: This class is characterized by intermittencies as illustrated in Fig.~\ref{fig-GPevolution2}. The dynamics can be interpreted in such a way that the trajectories jump irregularly between different coupling schemes. Accordingly, frequency locking between different pairs of modes is only established temporarily during the time evolution.

\begin{figure}[htb]
\begin{center}
\includegraphics[width=8cm,  angle=0]{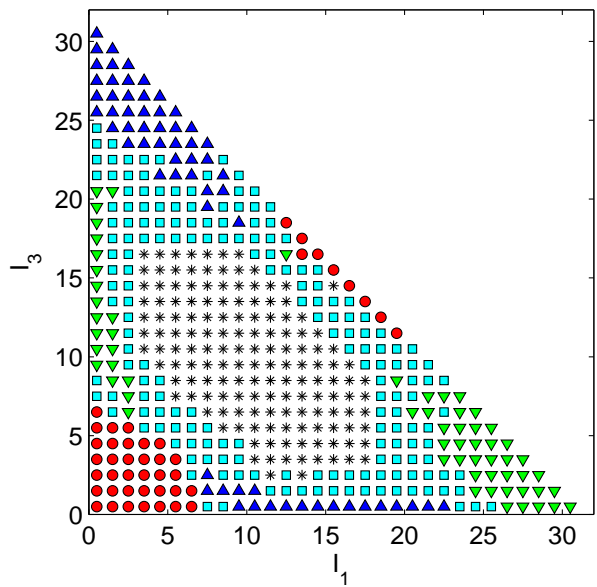}
\end{center}
\caption{\label{fig-basiscl}(Color online) Characterization of the classical actions $I_k=|c_k|^2$ through direct numerical integration of Eq.~(\ref{eq-GPE}). The action $I_2$ is given by $I_2=K-I_1-I_3$. Plotted are time evolutions of type (A) ($\circ$, red), type (B) ($\triangledown$, green), type (C) ($\triangle$, blue), type (E1) ($*$, black) and type (E2) ($\square$, cyan).}
\end{figure}

\begin{figure}[htb]
\begin{center}
\includegraphics[width=8cm]{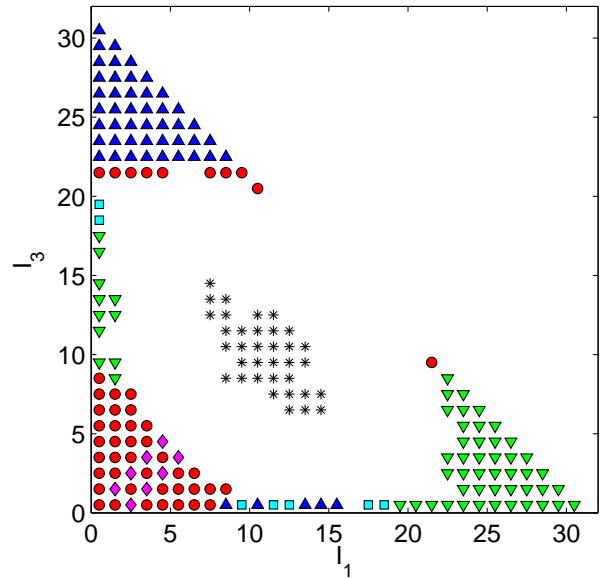}
\end{center}
\caption{\label{fig-basisqm}(Color online) Characterization of the classical actions $I_k=|c_k|^2$
through the semiclassical wave functions. The action $I_2$ is given by $I_2=K-I_1-I_3$. Plotted are 
actions whose quantum analog belongs to type (A) ($\circ$, red), type (B) ($\triangledown$, green), type (C) ($\triangle$, blue), type (D) ($\diamond$, magenta), type (E1) (*, black), and type (E2) ($\square$, cyan).}
\end{figure}

With this scheme, we can classify the dynamics of all possible basis states $\vec{b}(t)$, as shown in Fig.~\ref{fig-basiscl}. The interesting point is that we can compare these results with the information that we extract from the semiclassical wave functions. For this we compare for a given basis state $\ket{n_1,n_2,n_3}$ all eigenfunctions (\ref{eq-eigfred}) to which the basis state contributes significantly and assign a type (A)--(E2) to this basis state if possible. The result is shown in Fig.~\ref{fig-basisqm}.
The points with no symbol indicate states which cannot be assigned uniquely to a certain type. However, for the shown basis states one can see a close correspondence between the classical and the quantum system. Only at the fringes are there small deviations. 
Therefore the quantum mechanical analysis provides a grid of initial conditions for which we can predict the behavior of the solutions of the mean-field equations.
Finally, we remark, that the classification of the basis states in Fig.~\ref{fig-basiscl} holds in principle also for an arbitrary choice of the initial phases in Eq.~(\ref{eq-basisb}). Only at the fringes of the different zones does the behavior of the time dynamics depend crucially on the initial conditions and there it can deviate from this classification.
\subsection{Eigenstates}
In the last section we discussed the close resemblance between the quantum and the classical picture by assigning the same characterization scheme with types (A)--(E2) to the basis functions and the solutions of the mean-field equations. In this section we want to investigate, whether also the eigenstates of the quantum system can be reinterpreted classically, i.~e.\ if they can be used to identify the different types of dynamical behavior in the system of the three Bose-Einstein condensates weakly coupled by Josephson junctions.
We construct the classical analog of Eq.~(\ref{eq-eigf}) by defining the set of vectors 
\begin{equation}
\vec{B}(n_1,n_2,n_3) = (n_1+1/2,n_2+1/2,n_3+1/2)\,,
\end{equation}
which are related to the vectors $\vec{b}(t=0)$ by $B_k=b_k^2(0)$ (cf. Eq.~(\ref{eq-basisb})). However, note that the vectors $\vec{B}(n_1,n_2,n_3)$, like the vectors $\vec{b}(t;n_1,n_2,n_3)$ of Eq.~(\ref{eq-basisb}), do not form a basis of ${\mathbb C}^3$.
In analogy to Eq.~(\ref{eq-eigexpansion}) one can write
\begin{equation}
\vec{\ef}(t=0) = \!\!\!\!\!\!\!\sum_{n_1+n_2+n_3=N}\!\!\!\!\!\!\!
c_{n_1,n_2,n_3}^2\,\vec{B}(n_1,n_2,n_3)\,,
\end{equation}
where the real-valued coefficients $c_{n_1,n_2,n_3}$ are taken from Eq.~(\ref{eq-eigf}).
\begin{figure}[tb]
\begin{center}
\includegraphics[width=8cm,  angle=0]{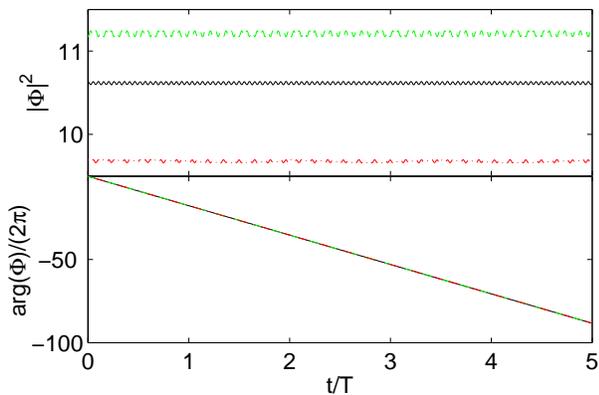}
\end{center}
\caption{\label{fig-GPeig1}(Color online) Time evolution of the mean-field equations for an initial condition corresponding to the first quantum eigenstate. Shown are squared modulus (top) and the phase of the first ($-$, black), second ($--$, green) and third ($-\cdot-$, red) mode. The time is measured with respect to $T=2\pi/\omega$.}
\end{figure}
In this naive approach, the vector $\vec{\ef}$
can be interpreted as the quantum expectation value of the action $\hat{\vec{I}}$
($\hat I_k = \hat n_k+1/2$) in the quantum state $\ket{\ef}$,
\begin{equation}
\scal{\ef}{\hat I_k|\ef} =  \sum_{n_1,n_2,n_3}c_{n_1,n_2,n_3}^2\,(n_k+{1}/{2})\,,
\end{equation}
where we have simply used the representation~(\ref{eq-eigexpansion}) of the eigenfunctions.
The initial phases are chosen equal zero like in the case of the basis vectors.
In order to use this vector $\vec{\ef}$ as initial conditions for the mean-field equations, we must take the square root of each component, and to this end we define the new vector $\vec{\phi}$ with components $\phi_k=\sqrt{\Phi_k}$. These vectors are normalized as  
\begin{equation}
|\vec\phi|^2=\sum_{k=1}^3\Phi_k= \!\!\!\!\!\!\!\sum_{n_1+n_2+n_3=N}\!\!\!\!\!\!\!
c_{n_1,n_2,n_3}^2\,\sum_{k=1}^3 B_k=K\,,
\end{equation}
where $K=31.5=N+3/2$ is the classically conserved total action of Eq.~(\ref{eq-Kcl}). In the context of the Gross-Pitaevskii equation, the norm of the condensate wave function gives the number of particles in the condensate. We get the additional term of $3/2$ for the number of particles compared to the many-particle Hamiltonian (\ref{eq-Hqm}), since we use the semiclassical correspondence of Eq.~(\ref{eq-Inrel1}). For Bose-Einstein condensates with a number of particles much larger than $30$, one can ignore the term $1/2$ in Eq.~(\ref{eq-Inrel1}) and obtain the standard correspondence between the particle numbers. However, for $N=30$, semiclassical studies like the present work show that the identification (\ref{eq-Inrel1}) gives a much better agreement between classical and quantum mechanics.
In order to obtain the normalization $|\vec{\phi'}|^2=1$, one simply has to set $\vec{\phi}=\vec{\phi'}\sqrt{N}$ and replace the nonlinearities $x_k$ by $x_k=g/K$.

In Fig.~\ref{fig-GPeig1}, the time evolution for the initial condition $\vec{\ef}_1$ is shown. The time evolution shows approximately constant occupations $|c_k|^2$ (upper panel), and the three phases are locked. In the reduced system, this corresponds to a point in the neighborhood of a fixed point. For the parameter values chosen in this article, there does not exist an exact fixed point of the Hamiltonian flow of the reduced system, although this point serves as guiding center for the wave functions of type (E1). In that sense the semiclassical wave functions behave very similarly in the neighborhood of a guiding center, while the solutions of the Gross-Pitaevskii equation are very sensitive to small deviations due to the nonlinearity of the time-evolution.

\begin{figure}[htb]
\begin{center}
\includegraphics[width=8cm,  angle=0]{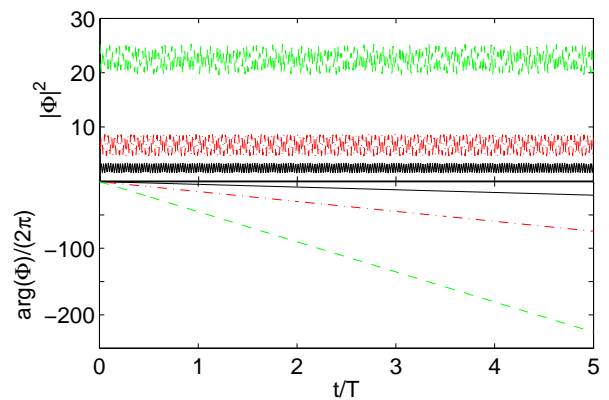}
\end{center}
\caption{\label{fig-GPeig444}(Color online) Time evolution of Eq.~(\ref{eq-GPE}) for the initial condition $\ef_{444}(0)$. Shown are squared modulus (top) and the phase of the first ($-$, black), second ($--$, green) and third ($-\cdot-$, red) mode. The time is measured with respect to $T=2\pi/\omega$.}
\end{figure}

Another example is shown in Fig.~\ref{fig-GPeig444} for a type (A) motion.
The phases of the modes evolve independently and the amplitudes show tiny oscillations, due to the fact that the time evolution does not coincide with the corresponding idealized guiding center of type (A). Because the system is dominated by the anharmonicities the effective frequencies are almost linear in the actions according to Eq.~(\ref{eq-omegaeff}). Therefore the slopes of the phase curves are proportional to the average values of the corresponding actions.

To conclude, from the classical point of view the
analysis of the corresponding quantum system offers a direct visual method for the understanding
of the structure and can be used to identify the dynamical behavior of the system of the three weakly coupled Bose-Einstein condensates in the mean-field approximation simply by diagonalizing the quantum Hamiltonian and plotting the eigenfunctions in the
appropriate basis.
\section{Conclusion}
In our investigation of a Bose-Einstein condensate in a multi-well potential, we showed a close correspondence between the quantum mechanical description and a classical version where the bosonic creation and annihilation operators of the many particle system are replaced by c-numbers. We truncated the many-particle Hamiltonian to a few relevant modes and obtained a system of three coupled anharmonic oscillators. Whether the truncation
at a small number of modes is justified depends crucially on an appropriate
choice of the expansion basis and on the external potential.
In order to compare the quantum system with its classical counterpart, we introduced the concept of the semiclassical wave functions defined on the same toroidal configuration space as in the classical system. This choice of the quantum mechanical representation allowed us to compare the quantum system directly with the classical system. In both cases, for the classical and the quantum system, we used the conserved particle number resp.\ total action to reduce the degrees of freedom to two. Classically, we can identify various geometric structures in phase space that are connected to different types of motion in the configuration space.  These different types of motion belonging to the various guiding centers, are also found in the quantum mechanical wave functions. So we used these guiding centers firstly to sort a large number of wave function into these different classes, and secondly to assign uniquely geometric quantum numbers to the wave functions within one class. In this geometric picture, the wave functions describe the quantum excitations of the underlying classical dynamics. As an application, we can  characterize the entanglement between the different modes and we can also determine the number of particles in each of the entangled modes using their associated quantum numbers.

In the last part of this article we analyzed the significance of the quantum mechanical classification of the wave functions for the classical dynamics. For this we studied classical trajectories which have initial conditions corresponding to quantum mechanical number states, or which correspond to the eigenstates directly. In both cases, we could obtain the characteristics of the semiclassical classification also from the classical trajectories, although the classical dynamics is much more sensitive to deviations from the idealized guiding centers.

Concluding, we showed that semiclassical wave functions provide an intuitive picture of the quantum mechanical many-particle eigenfunctions, and allow a direct classification of the dynamics.
\section*{Acknowledgments}
We thank H. S. Taylor for interesting discussions. Support by DGAPA under grant number IN-118005 is gratefully acknowledged. We thank the anonymous referee for an unusually detailed and careful referee
report which has helped us a lot to improve the final version of the 
manuscript. This work was supported by a fellowship within the Postdoc-Programme of the German Academic Exchange Service (DAAD).

\end{document}